\documentclass[aps,prl,twocolumn,groupedaddress,showpacs,amsmath,floatfix]{revtex4}
\usepackage{graphicx}

\newcommand{\Ket}[1]{\vert \, #1 \, \rangle}

\begin{document}


\title{Pulse calibration and non-adiabatic control of solid-state artificial atoms}

\author{Jonas Bylander$^1$} \email[bylander@mit.edu]{}
\author{Mark S. Rudner$^2$} \altaffiliation{Present addresses:  Harvard Physics Department, Cambridge, MA;}
\author{Andrey V. Shytov$^4$}
\author{Sergio O. Valenzuela$^3$} \altaffiliation{CIN2 ICN/CSIC, Bellaterra, Barcelona, Spain;}
\author{David M. Berns$^{1,2}$}
\author{Karl K.~Berggren$^5$} \altaffiliation{MIT EECS Department}
\author{Leonid S.~Levitov$^2$}
\author{William D.~Oliver$^{1,5}$}

\affiliation{$^1$Research Laboratory of Electronics,
$^2$Department of Physics,
$^3$Francis Bitter Magnet Laboratory,\\
Massachusetts Institute of Technology (MIT), Cambridge, MA 02139, USA,\\
$^4$Department of Physics, University of Utah, Salt Lake City, UT 84112, USA,\\
$^5$MIT Lincoln Laboratory, 244 Wood Street, Lexington, MA 02420, USA}

\date{\today}

\begin{abstract}

Transitions in an artificial atom, driven non-adiabatically through an energy-level avoided crossing, can be controlled by carefully engineering the driving protocol.
We have driven a superconducting persistent-current qubit with a large-amplitude, radio-frequency field.
By applying a bi-harmonic waveform generated by a digital source, we demonstrate a mapping between the amplitude and phase of the harmonics produced at the source and those received by the device.
This allows us to image the actual waveform at the device.
This information is used to engineer a desired time dependence, as confirmed by detailed comparison with simulation.
%
%
%
%
\end{abstract}



\maketitle

Due to the strong coupling of solid-state artificial atoms \cite{Clarke-Nature-08,Hanson-Nature-08} to electromagnetic fields, a wide range of pulsing techniques can be used to manipulate and control such systems.
A common approach, first developed for natural atomic systems,
involves low-frequency Rabi oscillations
driven by low-amplitude microwave 
radiation
\cite{Nakamura-PRL-01,Vion-Science-02-etal,Martinis-PRL-02,Chiorescu-Science-03}, combined with envelope-shaping techniques \cite{Steffen-PRB-03,Lucero-PRL-08-etal}.
%
However, strong driving affords many more possibilities for quantum control.
Fast dc pulses can be used to bring the system non-adiabatically into the vicinity of an avoided level-crossing, where coherent oscillations result from Larmor-type precession
\cite{Nakamura-Nature-99,Duty-PRB-04,Poletto-2008-etal}.

Furthermore, quantum coherence was recently shown to persist under large-amplitude harmonic driving
\cite{Izmalkov-Europhys-04-etal,Oliver-Science-05-etal,Sillanpaa-PRL-06,Wilson-PRL-07-etal,Berns-Nature-08-etal,Shytov-EurPhysJB-03,Ashhab-PRA-07}.
In this regime,
by carefully engineering the driving protocol, arbitrary rotations of a qubit's quantum state on the Bloch sphere could be performed.
Such protocols may lead to much faster quantum-logic gates than those achieved by using Rabi-transition based techniques.
To make such an approach feasible, one must be able to apply external fields of arbitrary time dependence to a quantum device.
With this ability, the techniques used for pulsed NMR can be extended to achieve even higher-fidelity quantum control \cite{Fortunato-JCP-02-etal}.


In addition to the challenge of designing optimized pulses, accurately delivering them to a device in a cryogenic environment is a difficult problem in its own right, especially at radio and microwave frequencies.
In particular with transient pulses \cite{Nakamura-Nature-99,Duty-PRB-04,Poletto-2008-etal}, it is hard to determine the exact pulse shape at the device.
Although digital waveform generators offer the possibility to create control pulses with essentially arbitrary time dependence, the signal that reaches the device may be strongly distorted due to the impedance mismatch and frequency dispersion that can occur in the long coaxial cables leading from the generator to the device.
Thus, in order to achieve high-fidelity quantum control, it is important to learn
not only how to design optimized pulses with arbitrary shape in the computer, but
also how to apply them faithfully to the device under study.

\begin{figure}
\centering
\includegraphics[height=1\columnwidth]{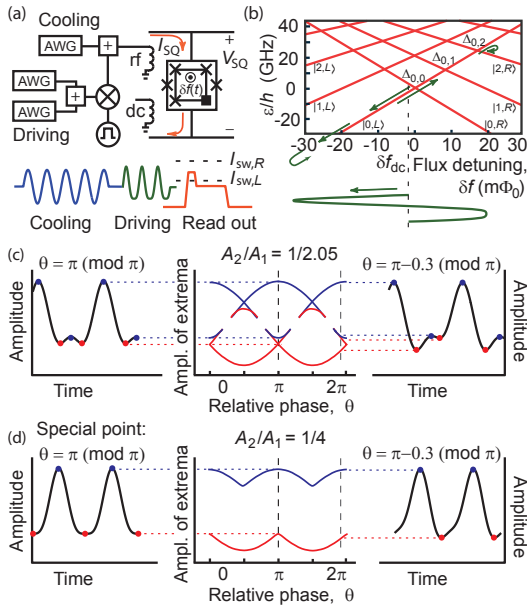}
\vspace{-4mm}
  \caption{\label{fig:diagrams} (color online)
  (a) Schematic experimental set-up.
  The device was placed 
  at 20\,mK (see  \cite{Oliver-Science-05-etal} for details).
  (b) Energy diagram determined by amplitude spectroscopy \cite{Berns-Nature-08-etal}. The energy gaps are small on the scale of the figure ($\Delta_{0,q}/h\!=\!0.013,$ 0.090, 0.40, and 2.2\,GHz for $q\!=\!0,\ldots, 3$, resp.).
  Below: flux excursion $\delta f(t)$ during the short, bi-harmonic pulse that probes the time-dependent dynamics in Fig.~\ref{fig:pwscan}.
  (c--d) Examples of bi-harmonic waveforms (\ref{eq:biharm_wf}). $\theta$-values are indicated by dashed, vertical lines; blue/red denote maxima/minima. The lower left waveform corresponds to the one in Fig.~\ref{fig:pwscan}(b).
}
\end{figure}

In this Letter we present methods to image the extremal points of a periodic signal and the actual waveform of a single pulse, as received by the device.
Applied to digitally generated, radio-frequency bi-harmonic pulses, Eq.~(\ref{eq:biharm_wf}), this approach allows us to determine the amplitude ratio and phase between harmonics, as illustrated in Figs~\ref{fig:diagrams} and \ref{fig:pwscan}.
Using this information, we create
waveforms that are calibrated to accurately produce the desired time dependence at the device.
In particular, the ability to engineer waveforms is used to control the time spent near an avoided crossing.
The effect of the latter on quantum evolution is illustrated and confirmed by detailed comparison with simulations.

In our experiments, we use a niobium superconducting persistent-current (PC) qubit, see Fig.~\ref{fig:diagrams}(a).
When the magnetic flux $f$ piercing the qubit loop is nearly $\Phi_0/2$, where $\Phi_0\!=\!h/2e$ is the flux quantum, 
the qubit's potential energy exhibits a double-well profile.
The system then supports a set of discrete, diabatic states $\{\Ket{p, L}, \Ket{q, R} ; \, (p, q)=0,1,\ldots \}$, localized in the left and right wells, and associated with opposing persistent currents.
Their energies 
vary linearly with the magnetic flux 
%
and exhibit avoided level crossings, see Fig.~\ref{fig:diagrams}(b). 

\begin{figure*}
\centering
\includegraphics[height=.5\columnwidth]{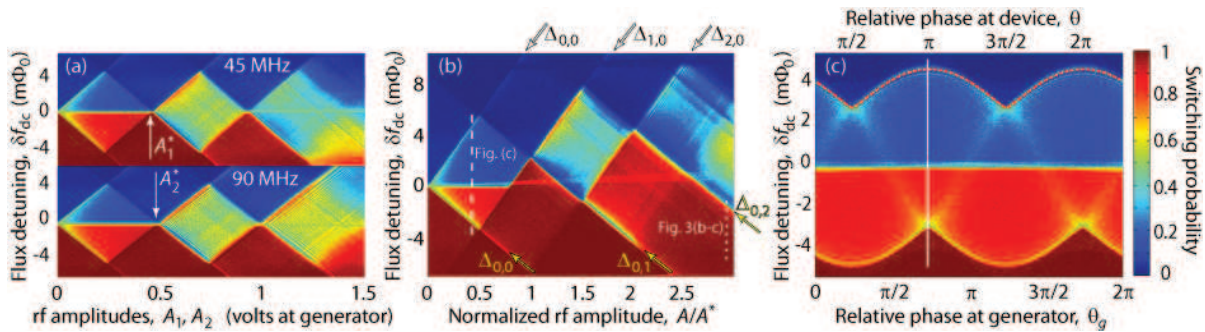}
\vspace{-4mm}
  \caption{\label{fig:diamonds} (color online)
  Calibration of rf amplitude and phase using long pulses.
  (a) ``Spectroscopic diamonds'' from single-harmonic driving.
  The points $A_{1,2}^*$ of optimal cooling (arrows) are used for amplitude calibration at each frequency.
  (b) Diamonds when driving with the bi-harmonic waveform $\{1/4,\, \pi\}$ at $\omega/2\pi\!=\!45$\,MHz, see Eq.~(\ref{eq:biharm_wf}).
  The arrows point out the diamond edges corresponding to the quartic extremum reaching the $\Delta_{0,\,q}$ crossings and the quadratic the $\Delta_{p,\,0}$.
  (c) Measurement along the dashed line in (b) for varying $\theta_g$. The quartic extremum $\theta\!=\!\pi$ is obtained at a cusp (solid white line), \emph{cf.\@} Fig.~\ref{fig:diagrams}(d).
}
\end{figure*}

\begin{figure*}
\includegraphics[height=.8\columnwidth]{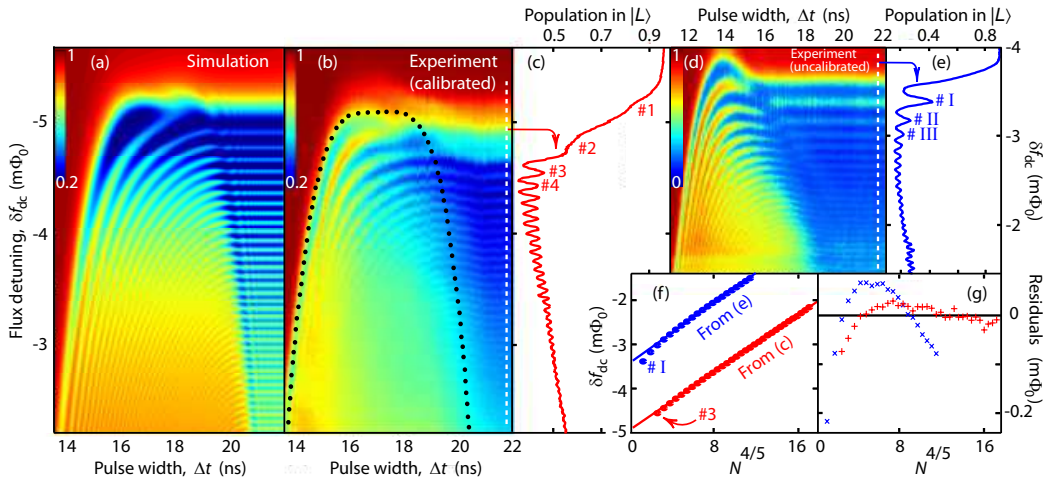}
\vspace{-4mm}
  \caption{\label{fig:pwscan} (color online) Time-dependent response.
   (a--b) Population in state $|L\rangle$ at the final time $\Delta t$ of a variable-length, flat-top pulse $\{1/4,\, \pi\}$, reaching through $\Delta_{0,2}$ as in Fig.~\ref{fig:diagrams}(b). The amplitude $A$ is indicated by the rightmost dotted line in Fig.~\ref{fig:diamonds}(b).
  The edges of the curved fringes provide a direct image of the pulse (black, dotted line), obtained experimentally by comparing (a) and (b). 
  (c) Population in state $|L\rangle$ after a return trip  ($\Delta t\!=\!22$\,ns) through $\Delta_{0,2}$.
  (d--e) Like (b--c) but with a waveform calibrated using the long pulses in Fig.~\ref{fig:diamonds}, thus failing to give the correct waveform.
  (f) The lower (red) and upper (blue) sets of data show the spacings of the St\"{u}ckelberg-interference fringes in (c) and (e), respectively, with fits to the $N^{4/5}$ power law (fringes  $N\!=\!4-\!35$ and $\mathrm{III}-\mathrm{XXI}$ fitted).
  (g) The residuals indicate that the fringes in (e) [blue $\times$] do not fit well, while those in (c) [red $+$] fit better; this confirms that the calibrated pulse better approximates the $\{1/4,\, \pi\}$ waveform.
}
\end{figure*}

We drive the system longitudinally with an applied flux $\delta f(t) \equiv f(t) - \Phi_0/2 = \delta f_{\rm dc} + f_{\rm rf} (t)$ consisting of a static and a time-dependent part.
This driving creates large-amplitude excursions through the energy-level diagram from the point of the static flux bias $\delta f_{\rm dc}$, as indicated by the arrows in Fig.~\ref{fig:diagrams}(b).
The quantum evolution is primarily adiabatic, except when the field sweeps through the energy-level avoided crossings $\Delta_{p,q}$ where Landau-Zener (LZ) transitions split the state into coherent superpositions of $\Ket{p, L}$ and $\Ket{q, R}$.
When the excursion extends over multiple levels, the interplay between LZ transitions at different crossings and intrawell relaxation ($\Ket{p', L} \rightarrow \Ket{p, L}$ and $\Ket{q', R} \rightarrow \Ket{q, R}$) gives rise to ``spectroscopy diamonds" in the saturated population \cite{Berns-Nature-08-etal} as shown in Fig.~\ref{fig:diamonds}(a).
Because the LZ-transition probability
$P_\mathrm{LZ}=1-\exp(-\pi\Delta^2_{p,q}/2\hbar v)$ is exponentially sensitive to the energy sweep rate $v$ through the avoided crossing, the magnitude of population transfer contains useful information about the driving signal in the vicinity of the avoided crossing.
The bright diamond edges occur for combinations of static flux detuning $\delta f_{\rm dc}$ and rf amplitude where an extremum of the driving waveform just reaches an avoided crossing. 
At these parameter values, 
the system spends the most time in the vicinity of the avoided crossing, thus allowing for maximum population transfer.
As described below, this phenomenon allows us to use the bright bands due to transitions at a particular avoided crossing to image the extrema of an rf pulse. 

In our experiment, we synthesized the control pulses digitally using the Tektronix AWG5014 arbitrary waveform generator with effectively 250 MHz analog output bandwidth.
The pulses were launched through a coaxial cable into an on-chip coplanar waveguide terminated in an ``antenna" (loop of wire). The frequency-dependent impedance of this antenna, as well as the resistive coaxial cable, makes our microwave line differ from an ideal 50-$\Omega$ system.
Moreover, its \emph{a priori} frequency-dependent, inductive coupling to the qubit calls for amplitude and phase calibration \emph{at the device} for each frequency component of the pulse.

To demonstrate our method of pulse imaging and calibration, we drive the qubit with the bi-harmonic signal
\begin{equation} \label{eq:biharm_wf}
  f_{\rm rf} (t) = A_1 \cos(\omega t \!+\! \theta) + A_2 \cos(2\omega t).
\end{equation}
We classify the waveforms, described by Eq.~(\ref{eq:biharm_wf}),
in terms of the relative amplitude ratio and phase between harmonic components using the notation $\{A_2/A_1,\,\theta\}$.
%
The waveforms can have either two or four extrema per cycle, depending on the values $A_2/A_1$ and $\theta$,
see Figs~\ref{fig:diagrams}(c--d).
%

The cross-over between the domains of two and four extrema per cycle occurs at $A_2/A_1 \!=\! 1/4$.
At this amplitude ratio, the phase $\theta \!=\! \pi$ is particularly interesting:
this waveform has one parabolic (quadratic) and one flat (quartic)
extremum,
the flatness of which arises from cancelation
of the quadratic terms in the Taylor expansions of the two harmonic components.
The presence of the flat extremum relies on the delicate balance of amplitudes and phases; hence we exploit the $\{1/4,\,\pi\}$-waveform to calibrate the relative
phase difference between the two frequency components at the device
as shown in Fig.~\ref{fig:diamonds}.

Each experiment uses the following pulse sequence, see Fig.~\ref{fig:diagrams}(a).
A static flux is applied using a superconducting coil, and maintained fixed at some value $\delta\!f_{\rm dc}$.
A harmonic, 11-MHz, 3-$\mu$s cooling pulse initializes the qubit to 
its ground state \cite{Valenzuela-Science-06-etal}.
After a short delay ($\sim\!\!20$\,ns), an rf-driving pulse of duration $\Delta t$ and amplitude $f_{\rm rf}$ is applied.
The mixer in Fig.~\ref{fig:diagrams}(a) is used to terminate the pulse after a time $\Delta t$ in the experiments shown in Figs~\ref{fig:pwscan}(b,d), but is not used 
in the experiments of Fig.~\ref{fig:diamonds}.
After another short delay, a sample-and-hold current pulse is applied to the hysteretic SQUID magnetometer \cite{SQUID-handbook}, which measures the qubit in the basis of the diabatic states, and is set to switch only if the qubit is in state $|L\rangle$.
The resulting voltage signal is amplified at room temperature and recorded with a threshold detector.
This sequence is repeated a few thousand times at a 
rate of 10\,kHz. 

First, we independently calibrate the amplitudes of the two components 1 and 2 by finding their points of optimal cooling \cite{Valenzuela-Science-06-etal} in a single-frequency spectroscopy diamond, see Fig.~\ref{fig:diamonds}(a).
These points occur when the flux excursion precisely reaches both opposite side-avoided
crossings $\Delta_{0,1}$ and $\Delta_{1,\,0}$ at zero dc bias. We denote these amplitudes by $A_{1,2}^*$, and define $A^*\!=\!A_1^*+A_2^*$ and $A\!=\!A_1+A_2$.
By adding the two harmonics together, fixing the amplitude at a value within the range $0\!<\!A/A^*\!<\!0.5$ (see Fig.~\ref{fig:diamonds}(b), dashed line), and scanning over flux detuning $\delta\!f_{\rm dc}$ and phase $\theta_g$ at the generator, we obtain Fig.~\ref{fig:diamonds}(c).
As mentioned above, high intensity occurs when an extremum of the waveform just
reaches an avoided crossing.
Thus the value of $\theta_g$ at the cusps in Fig.~\ref{fig:diamonds}(c) corresponds
to the desired value of $\theta\!=\!\pi$ at the device [{\it cf.} Fig.~\ref{fig:diagrams}(d)].
%

Applying the calibrated, composite $\{1/4,\,\pi\}$-waveform to the device, we obtained the bi-harmonic spectroscopy diamonds in Fig.~\ref{fig:diamonds}(b).
The diamonds are skewed compared to (a) due to the asymmetry of the signal $f_{\rm rf}(t)$ about zero; a quadratic extremum has a larger amplitude than a quartic.
Additionally, the data exhibit greatly enhanced
population transfer at the diamond edges arising from the flat extremum relative to that at the edges arising from the parabolic extremum; 
the same is seen at the cusps in Fig.~\ref{fig:diamonds}(c).
As discussed above, the reason is that the flat maximum makes the system spend more time at the avoided crossing than does the parabolic maximum.

The data in Fig.~\ref{fig:diamonds} reflect the averaged response of the qubit after a 3-$\mu$s pulse of base frequency $\omega/2\pi\!=\!45$\,MHz (135 cycles).
However, on short time scales, the pulse requires a few cycles to reach its steady-state amplitude, attributable to the build-up time of the standing-wave voltage at the imperfectly terminated end of the microwave line.
In order to adjust the pulse shape for short pulses, we investigate the brief, temporal Larmor-type oscillations induced by a single, non-adiabatic $\Delta_{0,\,2}$-passage, and the build-up of St\"{u}ckelberg oscillations occurring during the first return trip through that crossing.

When the system traverses the crossing, the LZ process creates a quantum superposition of the states associated with different wells.
Upon return, these two components interfere with the relative phase  \cite{Shytov-EurPhysJB-03,Oliver-Science-05-etal,Sillanpaa-PRL-06}
\begin{equation}\label{eq:accrued_phase}
\Delta\phi_{12} = \frac{1}{\hbar} \int_{t_1}^{t_2} \!\mathrm{d}t \,\Delta\varepsilon(t) ,
\end{equation}
where $\Delta \varepsilon(t)$ is the instantaneous diabatic energy-level separation. 
This leads to St\"{u}ckelberg-interference fringes in the occupation probabilities, provided that the time of the excursion, typically a fraction of the driving period,
is smaller than the relevant decoherence times \cite{Shytov-EurPhysJB-03}.

We apply a nominally flat-top pulse with an amplitude that reaches through $\Delta_{0,\,2}$.
At a time $\Delta t$, we abruptly terminate the pulse, and so obtain a snapshot recording of the population, see Fig.~\ref{fig:pwscan}(b).
Experimentally this is done in the following way.
A microwave cooling pulse initializes the system to the ground state $|0,L\rangle$ at $\delta\!f_{\rm dc}<0$.
Then, a bi-harmonic pulse (\ref{eq:biharm_wf}) starts the flux excursion towards the left, drawing the system deep into its ground state,
as shown by the arrows in Fig.~\ref{fig:diagrams}(b) (in this way we minimize the effects of unavoidable transients associated with the pulse turn-on).
%
Upon return towards the right, the system rapidly traverses the avoided crossings $\Delta_{0,\,0}$ and $\Delta_{0,1}$ before reaching $\Delta_{0,\,2}$
\footnote{Because $\Delta_{0,\,0},\,\Delta_{0,1} \ll \Delta_{0,\,2}$, the population change over one cycle is dominated by LZ transitions at $\Delta_{0,\,2}$.}.
We interrupt the excursion at the time $\Delta t$ by mixing the pulse with the falling edge of a square pulse, thus returning the flux detuning 
to the value $\delta\!f_{\rm dc}$.
The population can then be read out well before inter-well relaxation occurs.
Through this pulsing scheme, we can increment the pulse length $\Delta t$ in steps smaller than the AWG's sampling time.

For pulse lengths $\Delta t$ long enough for the system to complete a round trip through $\Delta_{0,\,2}$, 
the final populations in the left and right wells are stationary, \emph{i.e.\@} independent of $\Delta t$.
The spacings of the resulting horizontal St\"{u}ckelberg interference fringes in Fig.~\ref{fig:pwscan}(c) are determined by the phase (\ref{eq:accrued_phase}) gained over the course of the excursion beyond $\Delta_{0,2}$; as described below, these fringes thus constitute a signature of the part of the waveform extending beyond the avoided crossing.

The quantization condition for interference, $\Delta \phi_{12} \!=\! 2\pi N$
gives the static flux detuning $\delta f_{\rm dc}^{(N)}$ for the $N^\mathrm{th}$ node of constructive interference.
For generic pulses with extrema that can be fit to parabolae, Eq.~(\ref{eq:accrued_phase}) leads to an $N^{2/3}$ power law for the values of dc-flux detuning at the interference nodes \cite{Berns-Nature-08-etal}.
However, for the
$\{1/4,\,\pi\}$ flat-top waveform that we seek,
the quartic extremum gives rise to an $N^{4/5}$ power law,
which is the signature we use
to determine when the desired pulse has been attained.
In Figs~\ref{fig:pwscan}(f--g) we demonstrate 
the $N^{4/5}$ power law 
for a pulse that is calibrated by comparison to a simulation,
in which we numerically integrate the Schr\"{o}dinger equation for a two-level system with time-dependent driving (\ref{eq:biharm_wf}).
%
The initial state is taken to be the ground state of the system at detuning $\delta f_{\rm dc}$.
For further details, see \cite{Berns-Nature-08-etal}.
Overall there is a good agreement between simulation and experimental data.
%
We attribute the discrepancy around $\Delta t\!=\!18$\,ns, $\delta f_\mathrm{dc}\!=\!-5\,\mathrm{m}\Phi_0$ in Fig.~\ref{fig:pwscan}(b) to non-ideal pulse turn-off,
and the loss of fringe contrast 
to intrawell relaxation, digital noise, and phase jitter \cite{Berns-Nature-08-etal}.

%
%

We note that a rectangular-pulse waveform would ideally lead to a very different fringe pattern compared to the one in Fig.~\ref{fig:pwscan}, showing maximal contrast when the pulse height matches the level-crossing position in energy, and symmetrically decreasing at higher and lower pulse amplitudes.
Interestingly, in the pioneering work of Nakamura \emph{et al.\@} \cite{Nakamura-Nature-99}, where such pulses were used to drive a qubit, the observed fringe pattern had a strong asymmetry, overall resembling the data in Fig.~\ref{fig:pwscan}. This behavior can be attributed to a smooth turn-on and turn-off of the pulses used in that work.

In this work, we present an approach that in principle can be used to design an arbitrary waveform at the device, and use it to drive a superconducting qubit.
This technique, which is demonstrated with biharmonic driving, relies on calibrating the different Fourier harmonics comprising the waveform.
Having performed such a calibration, we are able to control the time spent by the qubit near an avoided crossing.
The details of qubit response are accounted for by comparison with  a simulation.




We gratefully acknowledge discussions with and support from T.\@ P.\@ Orlando.
%
We thank Y.\@ Nakamura, F.\@ Nori and T.\@ Yamamoto for helpful discussions; G.\@ Fitch,
V.\@ Bolkhovsky, P.\@ Murphy, E.\@ Macedo, D.\@ Baker, and T.\@ Weir
at MIT Lincoln Laboratory for technical assistance.
This work was sponsored by the U.S.\@ Government and the W.~M.\@ Keck
Foundation Center for Extreme Quantum Information Theory.
M.~R.\@ was supported by DOE CSGF, Grant No.\@ DE-FG02-97ER25308, and the NSF.
The work at Lincoln Laboratory was sponsored by the AFOSR under Air Force Contract No.\@ FA8721-05-C-0002.
%
%
Opinions, interpretations, conclusions and recommendations are those of the author(s) and are not necessarily endorsed by the U.S.\@ Government.

\vspace{-7mm}



\end{document}